\documentclass[useAMS,usenatbib]{mn2e}
\usepackage{graphics}
\usepackage{graphicx}
\usepackage{psfig}
\input{epsf}
\usepackage{amsmath}
\usepackage{amssymb}
\usepackage{url}

\title[CMB peak correlations from cosmic strings]{Peak-peak correlations in the cosmic background radiation from cosmic strings}
\author[M. Sadegh Movahed, B. Javanmardi \& Ravi K. Sheth]
{M. Sadegh Movahed $^{1,2,3}$, B. Javanmardi$^{1,2}$,  Ravi K. Sheth$^{3,4}$ \\
$^1$Department of Physics, Shahid Beheshti University, G.C., 
    Evin, Tehran 19839, Iran\\
$^2$School of Astronomy, Institute for Research in Fundamental Sciences (IPM), 
    P.O.Box 19395-5531, Tehran, Iran\\
$^3$The Abdus Salam International Centre for Theoretical Physics,
    Strada Costiera, 11,  Trieste 34151, Italy\\
$^4$Center for Particle Cosmology, University of Pennsylvania, 
    209 S. 33rd St. Philadelphia, PA 19104, USA}
\begin{document}
\maketitle

\begin{abstract}
We examine the two-point correlation function of local maxima in temperature fluctuations at the last scattering surface when this stochastic field is modified by the additional fluctuations produced by straight cosmic strings via the Kaiser-Stebbins effect.
We demonstrate that one can detect the imprint of cosmic strings with tension $G\mu \gtrsim 1.2 \times 10^{-8}$ on noiseless $1^\prime$ resolution cosmic microwave background (CMB) maps at $95\%$ confidence interval. Including the effects of foregrounds and anticipated systematic errors increases the lower bound to $G\mu \gtrsim 9.0\times 10^{-8}$ at $2\sigma$ confidence level.  Smearing by beams of order $4'$ degrades the bound further to $G\mu \gtrsim 1.6 \times 10^{-7}$.
Our results indicate that two-point statistics are more powerful than 1-point statistics (e.g. number counts) for identifying the non-Gaussianity in the CMB due to straight cosmic strings.
\end{abstract}


\section{Introduction}

The origins of the seeds of present large scale structures in the universe are still debated. It is believed that they are mainly primordial and produced some
time after Big-Bang. In this framework, there are two approaches:
 $(1)$ the freezing-in of quantum fluctuations of a scalar field during the so-called inflationary epoch \citep{guth81,lid93,stein95,lid99}; and/or
 $(2)$ topological defects as sources \citep{kib76,kib80}.  
Indeed topological defects can be formed during phase transitions between different vacuum states in an expanding universe, and cosmic strings are predicted by quantum field theory in cosmology
\citep{kib76,zeld80,vilinkin81,vak84,vil85,shell87,hind94,allen96,maxim99,vilinkin00,sak06,bevis08,matt09,bevis10}.
Both the inflationary and topological defects scenarios predict the same features for cosmic microwave background (CMB) power spectrum on large scales.  But at the intermediate and small scales, due to differences in the super-horizon scale behaviour of perturbations in these theories, the predictions are completely different.

The inflationary $\Lambda$ cold dark matter ($\Lambda$CDM) paradigm is consistent with today's high precision observations of the CMB.  Nevertheless, from both theoretical and observational points of view there are many motivations for other sources of anisotropies.  E.g., in hybrid inflation models, the brane-world paradigm and superstring theory, production of topological defects are crucial and inevitable
\citep{cope94,sak97,sar02,cop03,pog03,maj04,dva04,kib04,tye06}.
For more recent observational results see \citet{bevis08,hu12,ring12,kuro13}.

A cosmic string (CS) network which consists of infinite strings, loops and junctions of strings can generate gravitational waves as the universe evolves. 
Astrophysical evidence of CS depends on 
 1) the inter-commuting probability and
 2) the dimensionless string tension
      $G\mu$/$c^{2}\equiv \Lambda^2/M^2_{\rm {Planck}}$, 
      where $G$ is Newton's constant, $\mu$ is the mass per unit length 
      of the cosmic string and $\Lambda$ is the energy scale of the 
      string creation epoch \citep{vilinkin00,bevis08,bevis10}. 
We set $c=1$ throughout this paper. Determining bounds on the value of $\mu$ directly means limiting the basis of fundamental theories for CS production. In addition, observing cosmic strings not only is a kind of observational evidence for such theories, but also provides an opportunity to rule out or confirm theoretical models of particle physics.

To infer a reliable prediction and finding the observational signatures of cosmic strings, it is essential to understand the statistical properties of a typical cosmic string network. Since calculating the cosmic string components is still challenging topic, subsequently, all previous studies especially from simulation point of view have relied on one or more simplifications (for more details see \citep{bevis07,bevis08}). In order to examine the evolution of cosmic string network, the most important key is the Òscaling solutionÓ, which is consistent with the numerical simulations\citep{vilinkin81,kib85,Albrecht85,b861,b862,b31,b89,Albrecht89,b89,allen90,b90,shell90,austin93,martin96,vincent98,vanch05,martin06,vanch06,polch06,polch07,ring07,olum07,dub08,fra08,pill11,kuro13}. 
Several different groups have independently established various codes as well as theoretical methods to explore the evolution of cosmic string network and their imprints on the different observations such as CMB 
\citep{vilinkin81,kib85,Albrecht85,b861,b862,b31,b89,Albrecht89,b89,allen90,b90,shell90,austin93,martin96,vincent98,yama00,moore01,vanch05,martin06,vanch06,polch06,polch07,ring07,olum07,dub08,fra08,pill11}. Initial conditions and equations of motion have major roles for simulation of cosmic string network \citep{vak84,pill11}. During the evolution of cosmic strings, inter-commute phenomenon produces closed loops and makes kinks. These loops lose their energy through gravitational radiations or particles production, depending on their size. The probability of this inter-commutation in principle is not unity. Allen and Shellard used a high-resolution total-variation-non-increasing (TVNI) \citep{sod85} algorithm to simulate the behaviour of cosmic strings in an expanding universe during the radiation-dominated era \citep{allen90}. The same simulations but in matter-dominated epoch have been done in \citep{yama00,moore01}. They found that long strings have scaling behaviour and exhibit significant small-scale structure kinks and short wavelength propagating modes. In addition, loops produce peaks at scales very smaller than the horizon. On the other hands, the Abelian-Higgs model and one-dimensional effective theory- the Nambu-Goto action - are suitable to simulate cosmic string network\citep{moore98,vincent98,fra08}. The scaling behaviour, the correlation of cosmic string network and evaluation of dominant decay mechanisms in different epochs have been investigated in \citep{vilinkin81,kib85,b861,b862,b31,b89,Albrecht85,Albrecht89,allen90,austin93,martin96,vincent98,vanch05,vanch06,martin06,polch06,polch07,ring07,olum07,dub08,fra08,pill11}.  Historically, after many extensive numerical and theoretical studies concerning cosmic string network, the cosmological as well as astrophysical effects of cosmic strings have attracted  much attention. One of the most pioneering studies was done by \citep{bou88}. Since at the time of importance for CMB computations,  difference in scale is enormous, consequently, idealizations for cosmic string such as zero-width limit and unconnected segmentation of string network are hence justified \citep{kasu00,bevis07}.

There are some important ways to investigate the effect of cosmic string network on the temperature fluctuations at the last scattering surface: $(i)$ Nambu-Goto simulation of connected cosmic strings, $(ii)$ A model based on stochastic ensemble of unconnected segments  \citep{allen96,allen97,albr97,cont99,land04}  and $(iii)$ Abelian-Higgs model on a lattice and evolution of cosmic string network are evolved in terms of their corresponding fields \citep{kasu00,bevis07}.  
 Following previous approaches, the full Boltzmann equations are solved to compute the CMB fluctuations from cosmic string network and all relevant physics to first order are taken into account \citep{cou94,allen96,land03}. The forth way, so-called statistical approach, has been adopted by some groups \citep{pre93,pre93b,pre93a,jeo05,amsel08,brand080,brand081,movahed11}. In the latter approach, analytical and numerical models have been proposed to investigate the effect of cosmic strings on the CMB by means of nature of cosmic string. The main part of this approach is based on counting random multiple impulses, inflicted on photon trajectories by cosmic string network between the time of recombination and present era. It is well known that the direct implication based on the explicit recognition of discontinuity in the fluctuations of CMB is a unique signature of straight cosmic string, namely the Kaiser-Stebbins effect \citep{kaiser84,allen97,seljak97}. It has been demonstrated that to infer reliable results not only one should use robust methods but also a combination of powerful methods is necessary \citep{stark03}. Usually, in the first three approaches, the contribution of various components on the CMB map are modelled in the spherical or Fourier spaces, and then by integrating on the power spectrum the effect will be emerged.  In contrast, in the latter method, the effect of topological defects is modelled in the real space \citep{pre93,pre93b,pre93a,jeo05,amsel08,brand080,brand081} As explained in more details in \citep{brand080,movahed11}, since the statistical isotropy is valid as a major statistical property \citep{hajian03,hajian06,movahed111,planck1}, one can compute the TPCF at those scales, which have not been affected by boundary condition. The superposition of fluctuations produced by cosmic strings can generate new and extra peaks in temperature fluctuations, consequently finding such statistically meaningful footprints in the map in comparison with pure Gaussian signature including instrumental noise may potentially help us to get deep insight in the cosmic string detections and put constraints on the free parameters of cosmic string theoretical models. Finally, because of the phase coefficients in the Fourier analysis, it seems that many trivial imprints of cosmic strings diminish or at least are mixed with other observational phenomena; therefore, this is another motivation to investigate the imprint of cosmic strings in the real space. To this end we used the method introduced by \citep{pre93,pre93b,pre93a,jeo05,amsel08,brand080,brand081} in this paper.  To make the simulation more reliable, it is interesting to do following tasks: According to \citep{martin06}, the number of cosmic strings in radiation and matter era changes. So, it may be proper to improve our code to take this effect into account. The inter-commutation of straight strings to produce loops has been considered in \citep{pill11}  however in our current approach, one can ignore this contribution \citep{brand080,movahed11}. The junction of cosmic string is other interesting point that can be used to improve our results \citep{henry05,bevis080,urres08,b8}. Investigation of E-mode and B-mode polarizations are other topics to do \citep{pog06,bevis07}.


There are many constraints on the upper as well as lower values of cosmic string parameters from theoretical and observational perspectives. Pulsar timing and
photometry based on gravitational microlensing and gravitational waves require  $10^{-15}<G\mu<10^{-8}$ \citep{khlop82,khlop86,okn99,jenet,dam05,psh09,psh10,rich10}.  
The COSMOS survey requires $G\mu<3\times10^{-7}$ \citep{b5}.
The $21 cm$ signature of CS has been investigated in \citep{b6}. On the other paper by Hernandez \& Brandenberger, it has been demonstrated that cosmic string signal has  the overall  thermal
noise of an individual pixel in the Square Kilometre Array for string tensions $G\mu >2.5\times 10^{-8}$ \citep{Oscar12}. 
The LIGO and VIRGO collaborations have determined
$7\times10^{-9}<G\mu<1.5\times10^{-7}$ \citep{b7}.
In a recent paper, the stochastic gravitational waves from the European pulsar timing array place constrains $G\mu<5.3\times 10^{-7}$ \citep{sanidas12}. Based on probable global earthquake there is another lower limit on $G\mu$ which is down to ten orders of magnitude smaller than the cosmological events \citep{moto13}.

Another strong constraint on $G\mu$ comes from the investigation of temperature fluctuations at the last scattering surface \citep{pog99,sima00,land03,fra08}. The accumulation of anisotropies induced by CS on the fluctuations at last scattering surface can be divided into two categories: $1)$ anisotropies related to pre-recombination processes and created by the Kaiser-Stebbins effect \citep{kaiser84} and $2)$ the decay of string loops, which results in a stochastic background of gravitational waves \citep{fra08,kaiser84}. 
CMB analyses bound the cosmic string tension to be $G\mu< 6.4\times 10^{-7}$ \citep{kaiser84,pre93b,fra05,bevis05,wyman05,wyman06,bevis07,fra08,rich10}.
The effects of CS on the skewness of the one-point probability distribution of CMB temperatures \citep{b8},
 the TT power spectrum \citep{yamau10b}, 
 the B-mode polarization \citep{b9} 
have also been considered.  

Various detection methods have been explored, including:  
 Wavelet domain Bayesian denoising: $G\mu>6.3\times10^{-10}$ \citep{b10}. 
 The Canny algorithm: $G\mu>5.5\times10^{-8}$ \citep{brand081,brand080,brand09}.
 Level crossing analysis: $G\mu \gtrsim 4\times10^{-9}$ and $G\mu \gtrsim 5.8\times10^{-9}$ without and in the presence of instrumental noise, respectively
\citep{movahed11}. 

Recent observational constraints via WMAP and the South Pole Telescope yield $G\mu<1.7\times 10^{-7}$ at $95\%$ confidence interval \citep{hu12}. In the mentioned paper the aspect of polarization power spectrum to put robust constrains on the properties of CS have been discussed.  Other computations expressed $G\mu<0.7\times 10^{-6}$ with $f_{\ell=10}<0.11$ (the fractional contribution of cosmic string on the temperature power spectrum at $\ell=10$) \citep{bevis08}. The constraints on CS from future CMB polarization come from \citep{simon11}. The non-Gaussianinty imposed by CS has been considered by \citep{stark03,chi10,hobson98,hobson01}. The more complementary and recent considerations concerning non-Gaussianity due to CS can be found in \citep{hind09,hind10}. In mentioned pair papers bispectrum and trispectrum by CS generated by Nambu-Goto cosmic string simulations in the Friedmann-Lema\^{\i}tre-Robertson-Walker universe and those given by analytic calculation have been compated. 
I
Planck 2013 data, put the following conservative upper bounds due to post-recombination string contributions on the tension of cosmic string: Based on Nambu-Goto string model $G\mu<1.5\times 10^{-7}$ for $f_{10}<0.015$ and $G\mu<1.3\times 10^{-7}$ for $f_{10}<0.01$, also using abelian-Higgs model $G\mu<3.2\times10^{-7}$ for  $f_{10}<0.028$ \citep{planckstring}.

In the current study we concentrate on the discontinuities and fluctuations in the CMB map arising by CS from the Kaiser-Stebbins (KS) effect \citep{kaiser84}. This phenomenon can produce observational consequences on the anisotropies in the CMB with high degree of reliability in the high resolution map.

In what follows we study the two-point correlation function of local maxima or minima in the observed temperature maps \citep{Efstathiou87,ravi1,hevan01} to see if this is a useful probe of the extra roughness in the temperature distribution induced by strings.  
Previous work has shown that although the bispectrum of all pixels in a map at $5.5'$ resolution is not sensitive to a CS component, the two-point correlation function of local maxima is, especially on scales of order $10-20$ arc-minutes \citep{hevan01}.  The main goal of the present work is the capability of clustering approach to detect CS component and to quantify the limits of $G\mu$ which such a measurement can place based on Kaiser-Stebbins phenomenon.  

Section~2 describes how we generate mock maps of primordial Gaussian CMB, and how we incorporate the effects of straight cosmic strings.  In essence this is a straightforward combination of the algorithms in \citet{ravi1} with \citet{movahed11}.  We then identify peaks in these maps and study if the one- and two-point statistics of these peaks agree with the Gaussian prediction, quantifying the statistical significance of the differences.  
A final section summarizes why we conclude that 2-point statistics are much more efficient for identifying the presence of CS compared to 1-point statistics.


\section{Simulation and analysis of mock CMB maps}

This section describes how we simulate maps of the last scattering surface
\citep{pre93,pre93a,pre93b,brand080,brand081,brand09,movahed11}.  At first, our code creates pure Gaussian
fluctuations corresponding to the standard inflationary model with
$\Lambda$CDM components in a flat universe following \cite{Efstathiou87}. 
However, our program can be easily modified to other cosmological models 
for this purpose. Secondly, anisotropies produced by straight cosmic strings
by the Kaiser-Stebbins effect, from the last scattering surface up to 
the present, are simulated following \cite{pre93b} and~\cite{brand080,brand081,brand09,movahed11}. 
Our method of simulation differs from that used to produce the CS maps 
analysed by \citep{hevan01}, where the CS contribution to the 
energy-momentum tensor was modelled using Fourier methods \citep{pog99}.

For reasons discussed in detail in \cite{brand080,brand081,brand09,movahed11}, to simulate the Kaiser-Stebbins lensing due to moving cosmic strings we work in real space, where straight strings produce random jumps on the background radiation field. The scaling behaviour of straight strings means that the number of strings crossing a given Hubble volume is fixed to $M_{\rm string}=10$ \citep{b31}.  The cosmic strings possess relativistic velocities; consequently, after $2t_{\rm H}$ ($t_{\rm H}$ is the Hubble time) an entirely new network of cosmic strings provides new kicks to the CMB photons. 
The two signals are superposed, then smeared by our model of the 
instrumental beam, after which we add instrumental noise.  
Finally, we identify peaks in the simulated maps and measure the peak-peak 
correlation, showing results after averaging over a large ensemble of 
realizations.

\begin{figure}
\begin{center}
\includegraphics[scale=0.4]{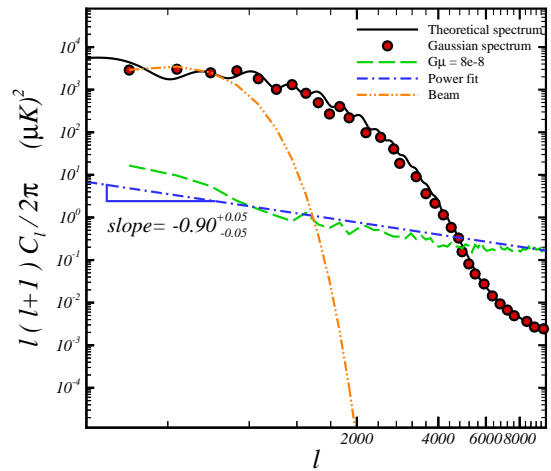}
\caption{Contribution of various components to the total angular power spectrum of temperature fluctuations. Solid line shows the power-spectrum for a WMAP-7 $\Lambda$CDM model (from CAMB).  Symbols show a simulated Gaussian map with resolution $R=1'$ and size $10^{\circ}$.   
Long-dashed line shows the contribution from cosmic strings having $G\mu=8\times10^{-8}$; dot-dashed line shows a power-law $\ell(\ell+1)C_{\ell}\sim\ell^{-0.90^{+0.05}_{-0.05}}$.  Dashed-dot-dot curve shows the window associated with a beam having FWHM$=20'$.  Clearly, the CS component is most easily detected at $\ell\sim 6000$.}\label{power1}
\end{center}
\end{figure}

\subsection{Mock Gaussian CMB map}
In what follows, the size and resolution of simulated map are 
$\Theta$ and $R$, respectively.  Thus, a $\Theta=10^\circ$ map at 
$R=1'$ requires $600\times 600$ pixels.  The rms instrumental noise, 
$\sigma_{noise}$, and the full width half maximum of detector, 
${\rm FWHM}$, are used to take into account additional effects 
on the simulated maps.  We set $\sigma_{noise} =10\mu K$ which is 
appropriate for the South Pole Telescope \citep{SPT,SPT11}.  We also use ${\rm FWHM} = 4'$ to illustrate our results.

For making Gaussian maps, all that is required is the initial 
power spectrum $C_\ell$.  For this, we use the CAMB software \citep{b29} 
with parameters appropriate for a $\Lambda$CDM model that is 
consistent with the WMAP-$7$, Supernova type Ia (SNIa) 
and the Sloan Digital Sky Survey (SDSS) datasets.  We use this 
$C_\ell $ to generate a 2D Gaussian random field following 
\cite{Efstathiou87}.  Since we are interested in relatively small 
angular scales, we work in the flat sky approximation, following \cite{ravi1}.

To add the effects of strings we follow \citep{brand080,brand081,brand09,movahed11}.
This means that we ignore the contribution of CS loops, since their size is smaller than our map resolution ($1$ arcmin).  In contrast the characteristic length scale of straight CSs is the Horizon scale.  

\begin{figure}
\begin{center}
\includegraphics[scale=0.43]{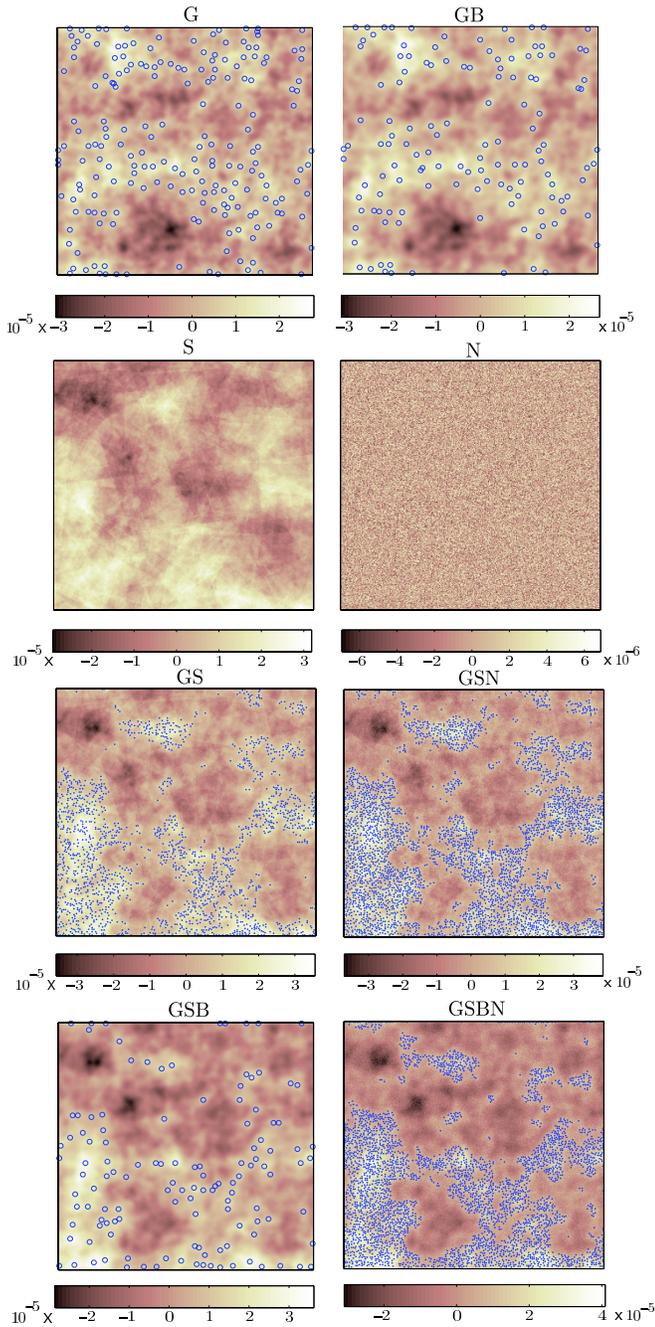}
\caption {\small Various components of simulated maps, for a WMAP-7 $\Lambda$CDM model, with $G\mu=8\times 10^{-7}$ strings.  The map size is $5^{\circ}\times 5^{\circ}$ at resolution $R=1'$ and smeared by a beam of FWHM=$4'$.  Circles in some panels show peaks above $\vartheta=0.5\sigma_0$.  }\label{map2}
\end{center}
\end{figure}

\subsection{Combination of simulated components}
Before starting to combine different components, for the sake of convenience, let us take note of the notation used. Throughout this paper we use $G$ for Gaussian component, $S$ for string part, $B$ stands for beam effect and $N$ is for noise part. In addition for combined map, Gaussian+String+Beam+Noise, we use $GSBN$.  
When combining the Gaussian ({\it G}) and string ({\it S}) components, we are careful to ensure that, at $\ell<200$ \citep{brand081,movahed11}, the total power is the same as that observed.  In practice, this means that in each pixel $(i,j)$ we set the fluctuation, ${\mathcal F}\equiv (T(i,j)-\langle T \rangle)/\langle T \rangle$, to be 
\begin{equation}\label{combination}
 {\mathcal F}_{(G+S)}(i,j) =  \omega {\mathcal F}_{(G)} (i,j)+ {\mathcal F}_{(S)}(i,j)
\end{equation}
where $\omega < 1$ is chosen so that the amplitude of the power spectrum 
\begin{equation}
 C_\ell^{(G+S)}=\omega^{2} C_\ell^{(G)} + C_\ell^{(S)}
\end{equation}
is close to that observed at $\ell<200$.  Since $C_\ell^{(S)}$ depends on $\ell$, determination of the appropriate value of $\omega$ is done by a likelihood analysis. 

The beam smearing is modelled by:
\begin{equation}
 C_\ell^{(G+S)B}=C_\ell^{(G+S)}\,{\rm e}^{-\Gamma^{2}\ell(\ell+1)}
\end{equation}
with $\Gamma={\rm {FWHM}}/\sqrt{8\ln2}$ \citep{Efstathiou87,ravi1}. 

Finally, we add a model for the noise:
\begin{equation}
\mathcal{F}(i,j)\equiv \mathcal{F}_{(G+S)B}(i,j)+\mathcal{F}_{(N)}(i,j),
\end{equation}
where the final noise term is white, i.e., it has $\langle \mathcal{F}_{(N)}(\textbf{r}_1)\mathcal{F}_{(N)}(\textbf{r}_2)\rangle\sim \delta_{\rm{Dirac}}(\textbf{r}_1-\textbf{r}_2)$, with the noise in pixel $(i,j)$ being a zero-mean Gaussian number with rms $\sigma_{noise}$.  Figure \ref{power1}  illustrates the comparison of power spectrum of Gaussian simulated map and that of given by observation (here observation is that of computed by e.g. CAMB software).  Figure~\ref{map2} illustrates various components and steps in our map making process.

\begin{figure}  
\begin{center}
\includegraphics[scale=0.45]{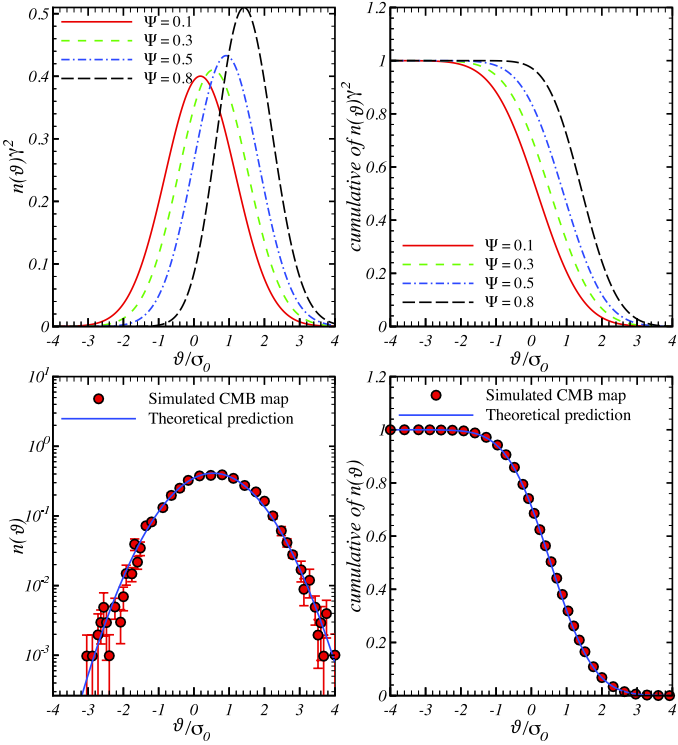}
\caption {\small Left upper panel indicates the normalized number density of peaks in a typical 2D Gaussian field while the right upper  panel shows the normalized cumulative peaks for the same map as a function of ${\vartheta}$ for various values of $\Psi$. The lower panels correspond the same just for our simulated Gaussian CMB map (size= $10^{\circ}$ and resolution is $R=1'$) by taking ensemble averages with $\Psi=0.320\pm0.020$ and $\gamma=(1.050\pm0.040)\times 10^{-3}$ at $1\sigma$ confidence interval. }\label{numberdensity2}
\end{center}
\end{figure}
\begin{figure}
\includegraphics[scale=0.4]{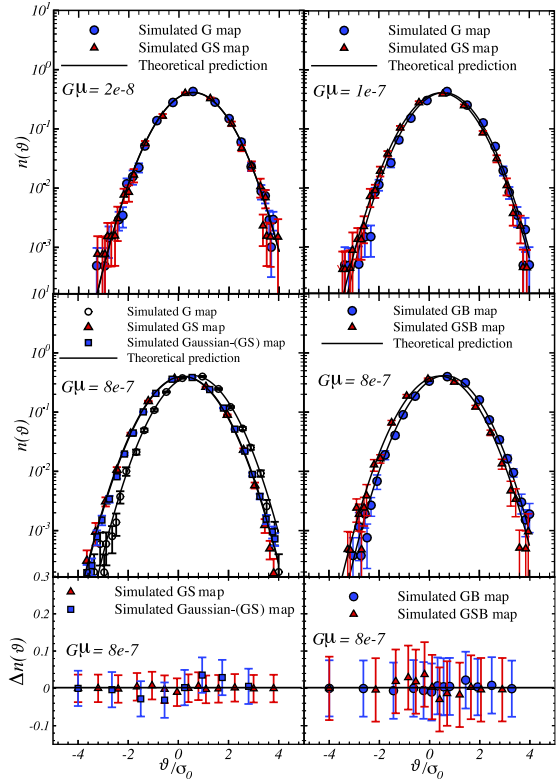} \\
\caption {\small Abundance of density of peaks as a function of peak threshold level for pure Gaussian maps and upon adding a cosmic string component. Upper left: $G\mu=2\times 10^{-8}$. Upper right: $G\mu=1\times 10^{-7}$. Middle left: $G\mu=8\times 10^{-7}$. Results for a Gaussian map (Gaussian-(GS)) having the same power spectrum as the Gaussian+String map has also been indicated in this panel. Middle right:  for $G\mu=8\times 10^{-7}$  with beam effect.  The lower panels show the difference between theoretical Gaussian field prediction for the number density of peak and what measured in the simulated maps. }\label{numberdensity}
\end{figure}

\subsection{Peak counts in mock maps}
We have checked that the number density of peaks we identify in the Gaussian maps agrees with that expected from theory.  When the peak height is expressed in units of the rms temperature, $\vartheta = \Delta T/T$, 
this prediction depends only on the shape of the power spectrum $C_\ell$ \citep{Efstathiou87}.  Since CS modify $C_\ell$ at high $\ell$, it is interesting to ask if the peak counts predicted by $C_\ell^{(G+S)}$ provide a good description of the peak abundances in the $G+S$ maps, even though the maps themselves are not Gaussian.  If not, then peak counts alone allow one to distinguish between a purely Gaussian model and one with an additional component.  
To compute theoretically the number density of extrema of a typical 2D Gaussian stochastic field in the flat sky approximation, we should construct the multivariate distribution function of underlying filed.  The components of mentioned distribution function are as $\vec{\mathcal{W}}= \{ \mathcal{F}, \mathcal{F}_{x},
\mathcal{F}_{y},\mathcal{F}_{xx},\mathcal{F}_{xy},\mathcal{F}_{yy}
\}$. Here $\mathcal{F}_{\alpha\beta}\equiv
\frac{\partial^{2}\mathcal{F}}{\partial {\alpha}\partial{\beta}}$. Consequently, according to notation introduced by Rice \citep{rice54} and \citep{peac85}, the multivariate distribution function of $6$ variables for a Gaussian process reads as:
\begin{equation}\label{JPDF11}
p(\vec{\mathcal{W}})=\sqrt{\frac{1}{(2\pi)^{6}{\rm det} \mathcal{M}}} \
e^{-\frac{1}{2}(\mathcal{W}^{T}.\mathcal{M}^{-1}.\mathcal{W})}
\end{equation}
where $\mathcal{M}$ is the covariance matrix of underlying variables, namely ${\mathcal M}=\langle {\mathcal W} \otimes{\mathcal W}\rangle $. Therefore number density of extrema for a purely Gaussian CMB map in the range of $[\vartheta,\vartheta+d\vartheta]$ is $n(\vartheta)d\vartheta$ and is given by:
\begin{equation}\label{nu11}
n({\vartheta})=\int p(\vec{\mathcal{W}}|\mathcal{F}_{\alpha}=0)\left |{\rm det}\mathcal{F}_{\alpha \beta}\right| d\vec{\mathcal{W}}
\end{equation}
Above expression can be integrated analytically in the Gaussian case and becomes \citep{bardeen86,Efstathiou87}:
\begin{equation}\label{nu1}
n({\vartheta})=\frac{1}{(2\pi)^{3/2}\gamma^2}e^{-{\vartheta}^2/2}{\mathcal{G}}({\Psi},{\Psi}{\vartheta})
\end{equation}
where
\begin{eqnarray}\label{g1}
&&{\mathcal{G}}({\Psi},{\Psi}{\vartheta})\equiv ({\Psi}^2{\vartheta}^2-\Psi^2)\left\{1-\frac{1}{2}{\rm erfc} \left[ \frac{{\Psi}{\vartheta}}{\sqrt{2(1-\Psi^2)}}\right]\right\}\nonumber\\
&&+{\Psi}{\vartheta}(1-\Psi^2)\frac{e^{-\frac{{\Psi}^2{\vartheta}^2}{2(1-\Psi^2)}}}{\sqrt{2\pi(1-\Psi^2)}}\nonumber\\
&&+\frac{e^{-\frac{{\Psi}^2{\vartheta}^2}{3-2\Psi^2}}}{\sqrt{3-2\Psi^2}}\left\{ 1-\frac{1}{2}{\rm erfc}\left[\frac{{\Psi}{\vartheta}}{\sqrt{2(1-\Psi^2)(3-2\Psi^2)}}\right]\right\}\nonumber\\
\end{eqnarray}
in which $\rm erfc(:)$ stands for complementary error function. According to notation explained in ref. \citep{Efstathiou87}, the so-called spectral parameters $\Psi$ and $\gamma$ in Eqs. (\ref{nu1}) and (\ref{g1}) are defined by:
$\Psi\equiv \frac{\sigma_1^2}{\sigma_0\sigma_2}$ and $\gamma\equiv\sqrt{2}\frac{\sigma_1}{\sigma_2}$. Where

\begin{eqnarray}\label{moments1}
&&\sigma_0^2\equiv\left \langle\mathcal{F}(\textbf{r})^2\right\rangle=\frac{1}{(2\pi)^2}\int S(|\textbf{k}|)d\textbf{k}\nonumber \\
&&\sigma_n^2\equiv \left\langle\left(\frac{\partial^n \mathcal{F}(\textbf{r})}{\partial x^n}\right)^2\right\rangle=\frac{1}{(2\pi)^2}\int k^{2n}S(|\textbf{k}|)d\textbf{k}
\end{eqnarray}
and $S(|\textbf{k}|)$ is spectral density. The number density of peaks in a 2D Gaussian map as a function of ${\vartheta}$ has been plotted in the upper panel of Fig. \ref{numberdensity2} for various values of spectral parameters.  We also compute the number density of our simulated pure Gaussian CMB map and illustrated in the lower panel of Fig. \ref{numberdensity2}.

Fig.~\ref{numberdensity} shows that, for $G\mu\le 10^{-7}$ (top panels), the peak counts in $G$ and $G+S$ are almost indistinguishable.  For larger values of $G\mu$, the peak counts are noticably different from one another (middle left panel), with the distribution being shifted to smaller mean values when CS are present.  However, the measurements are each well described by Gaussian peaks theory (solid line) \citep{Efstathiou87} with their respective power spectra ($C_\ell^{(G)}$ for the circles, and $C_\ell^{(G+S)}$ for the triangles), even though the $G+S$ maps themselves are not Gaussian.  Thus, given only the observed $C_\ell$ and the peak counts, it will not be possible to determine\ref{numberdensity} shows $\Delta n(\vartheta)\equiv n_{com.}(\vartheta)-n_{the.}(\vartheta)$. Where "{\it com.}" refers to numerical result and "{\it the.}" corresponds to theoretical prediction. 
\begin{figure}
\includegraphics[scale=0.45]{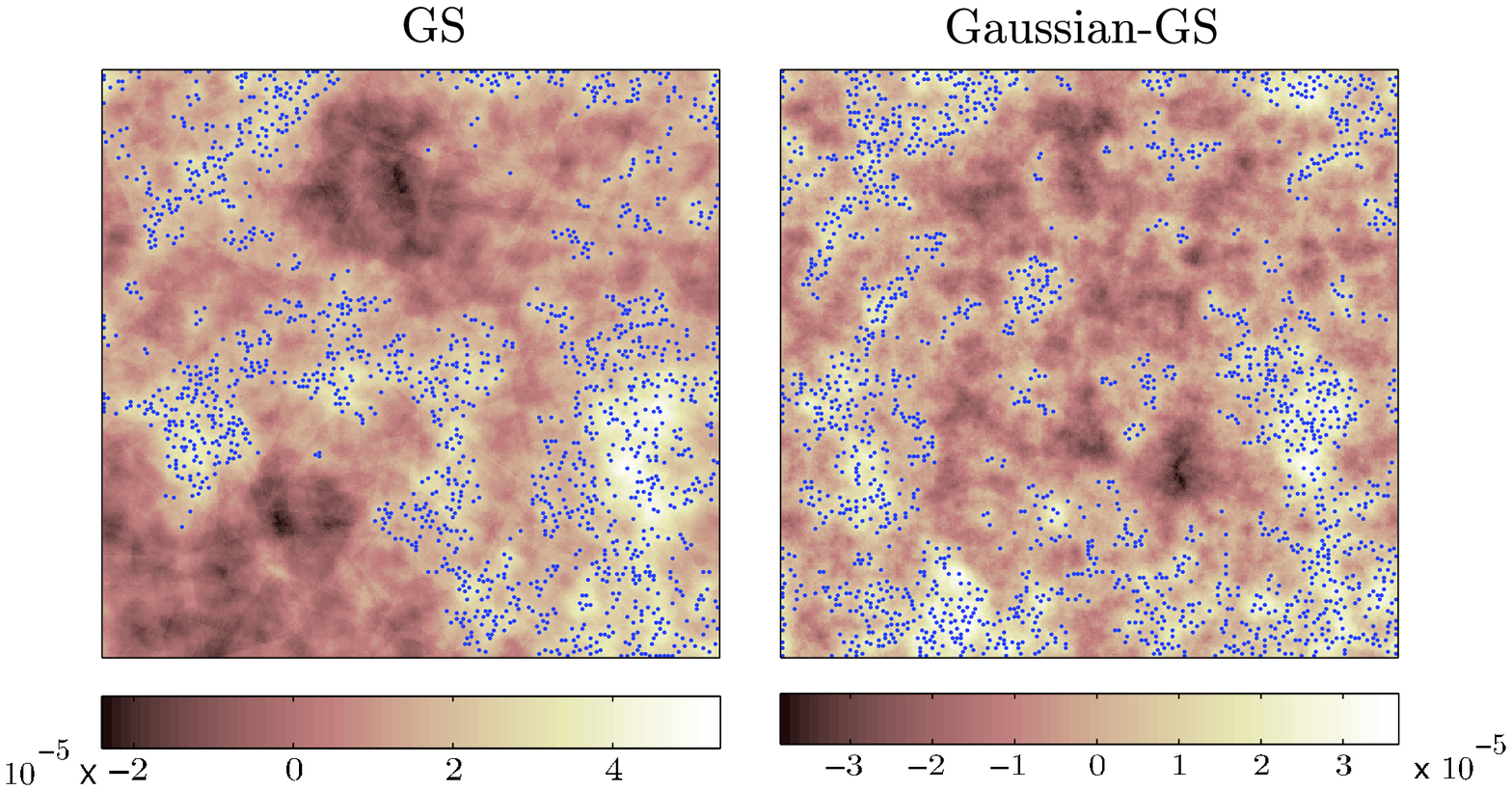} \\
\includegraphics[scale=0.45]{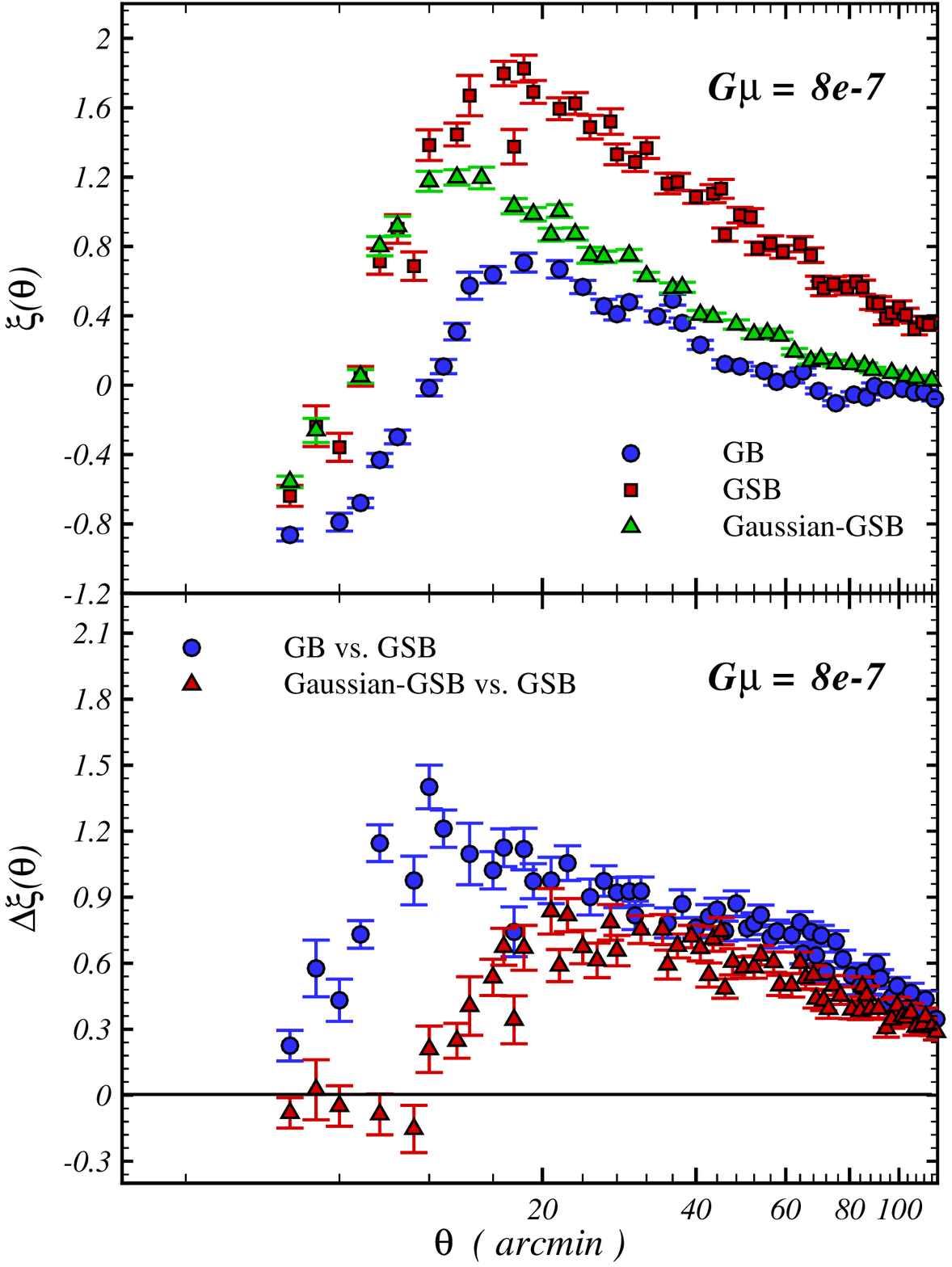} \\
\caption {Upper panel: Comparison of a Gaussian+String map (left) with a pure Gaussian map which has the same total power spectrum (right); blue dots show peaks above $\vartheta = 0.5\sigma_0$.  In both cases, $G\mu=8\times 10^{-7}$, the resolution $R=1'$ and the map size shown is $5^{\circ}\times 5^{\circ}$.  The morphology of these two maps is quite different. Lower panels:  TPCF of $\vartheta= 1.0\sigma_0$ peaks in these maps and differences between them. }\label{gs-gaussian}
\end{figure}

Recently,  \cite{pogo11} derived expressions for the number density of extrema in weakly non-Gaussian 2-Dimensional fields. They showed that various non-Gaussian models could be distinguished by means of $n(\vartheta)$. Our analysis demonstrates that, at least for the non-Gaussianity due to straight CSs, this does not work in our analysis.
In addition,  \cite{Rossi11} used excursion sets, such as regions above or below a temperature threshold and their clustering to examine the contribution of primordial non-Gaussianity on the CMB map. They also noticed to the optimum value of threshold, namely $\vartheta=2.00\sigma_0$ for discrimination between Gaussian and local non-Gaussian case. According to lower panel of  Fig. \ref{numberdensity}, it is evident that in examining cosmic string, the sensitivity of number density of peaks for wide range of thresholds is flat, so there is no priority in selecting the value of threshold for further computation.    
\begin{figure}
\includegraphics[scale=0.5]{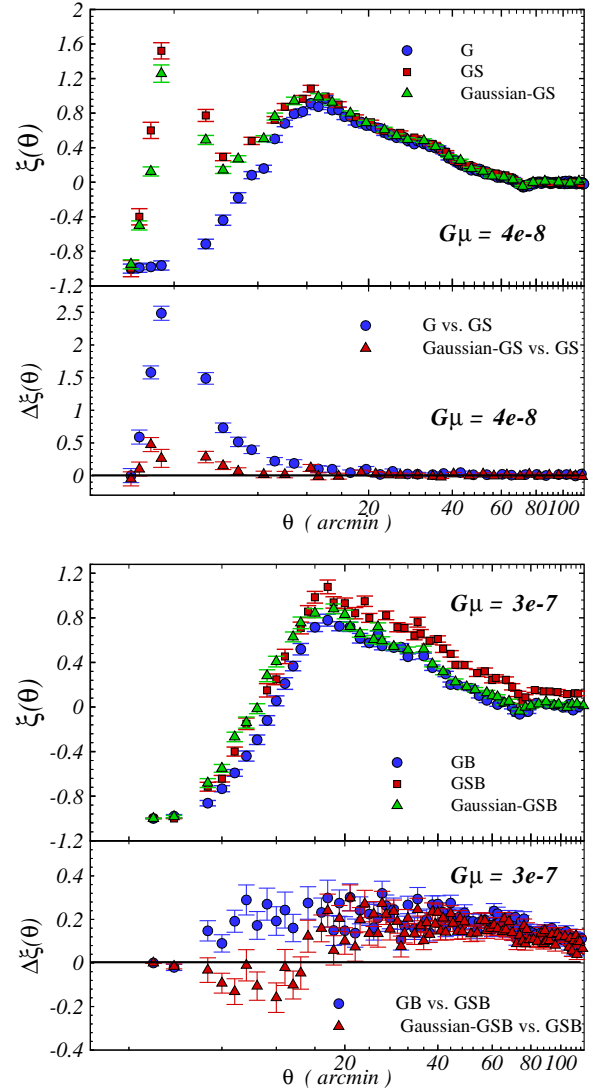}\\
\caption {\small Two-point correlation function of peaks above $\vartheta=1\sigma_0$ for different values of $G\mu$. Here FWHM is $4'$. $\Delta \xi (\theta)$ corresponds to difference between TPCF of various cases indicated in plots.}\label{two22}
\end{figure}

\subsection{Two-point statistics}
Although we have demonstrated that peak counts in our $G+S$ maps are consistent with those in a Gaussian field having the same $C_\ell$, direct inspection of the maps themselves (top panel of Figure~ \ref{gs-gaussian}) shows that they have quite different morphologies.  The CS component seems to add small scale random noise on top of the original Gaussian CMB signal.  We turn therefore to the use of two-point peak statistics for distinguising between the two maps.  

To this end, we measure the TPCF of peaks in our Gaussian maps, our $G+S$ maps, and our Gaussian-GS maps.  For each value of $G\mu$, map size, resolution scale and beam size, we have generated ensembles of $\sim 100$ maps. Lower panel of Figure~\ref{gs-gaussian} and Figure~\ref{two22} show results from averaging over 100 realizations of maps with $\Theta=10^\circ$ map at $R=1'$ and $\vartheta > 1\sigma_0$. It must point out that, the plots do not show the entire range of scales we simulated, only those that we believe to be accurate, free of cosmic variance and boundary effect. There are obvious differences between the TPCF in the $G$ and $G+S$ maps, with the latter having substantially more signal on small scales.  Although the beam erases some of this (Figures~\ref{gs-gaussian} and \ref{two22}), a residual effect remains.  This signal is rather different from that measured in a Gaussian field which has the same $C_\ell$ (what we called Gaussian-GS previously).  So we conclude that this is indeed a promising method for identifying the CS component in the maps.  

The lower panel of Figure~\ref{gs-gaussian} shows explicitly that, although the peak counts were unable to distinguish between $GSB$ and Gaussian-$GSB$ maps (Figure~\ref{numberdensity}), the TPCF on scales $\theta\ge 12'$ can.  Figure~\ref{two22} shows that the ability to discriminate depends on $G\mu$ and the beam size FWHM. 

It has been demonstrated that long strings exhibit significant small-scale kinks and short wavelength propagating modes  \citep{b89,Albrecht89,allen90,b90,shell90,vilinkin00,yama00,moore01}. In addition, we are interested in the value of $G\mu$ in the range of less than almost $2\times 10^{-7}$. If $f_{10} \sim 0.1$ then strings should dominate for $\ell>3000$ and remain above thermal Sunyaev-ZelÕdovich effect and for $\ell<1000$ with an acceptable $G\mu$ the effect of cosmic strings and texture are ignorable \citep{kaiser84,fra05,allen97,seljak97}. In addition, according to Fig. \ref{power1} and paper by \citep{fra08}, the power spectrum of cosmic strings behave as $\ell^{-0.9}$ (Fig. \ref{power1}), consequently, the contribution of cosmic strings for large mode, namely $200<\ell<1000$ is less than $1-2$ orders of magnitude, in addition since in this work we are working on local sky map instead full sky map namely $\ell_{min}>51$ for map size $10^{\circ}$ and $\ell_{min}>102$ for map size $5^{\circ}$, consequently we don't expect that cosmic strings have effective role in the large modes. To make the footprint of cosmic string in the peak-peak correlation function of CMB  more obvious, we took almost large value of $G\mu=8\times 10^{-7}$ in the lower panel of Fig. \ref{gs-gaussian}. We also checked the consistency between number density of peaks computing directly from our simulation and that of given by theoretical prediction (see Fig. \ref{numberdensity2}).

\subsection{Quantitative limits}
To quantify this we first compute the Student's t-test based on:
\begin{equation}\label{t-test}
t(\theta)=\frac{\xi_{(\diamond)}(\theta)-\xi_{(\otimes)}(\theta)}{\sqrt{\sigma_{(\diamond)}^{2}(\theta)+\sigma_{(\otimes)}^{2}}(\theta)}
\end{equation}
where $\xi(\theta)$ is the TPCF and $\sigma(\theta)$ is the mean standard deviation of each term in the numerator. The symbols $\diamond$ and $\otimes$ correspond to the $G+S$ and $G$ measurements and to $(G+S)B$ and $(G)B$ with beam effect. For each $\theta$, the corresponding P-value, $p(\theta)$, are calculated. Degrees of freedom based on the $t$-distribution function are $2 N_{sim}-2$, where $N_{sim}$ is the number of simulated maps. 

\begin{figure}
\includegraphics[scale=0.4]{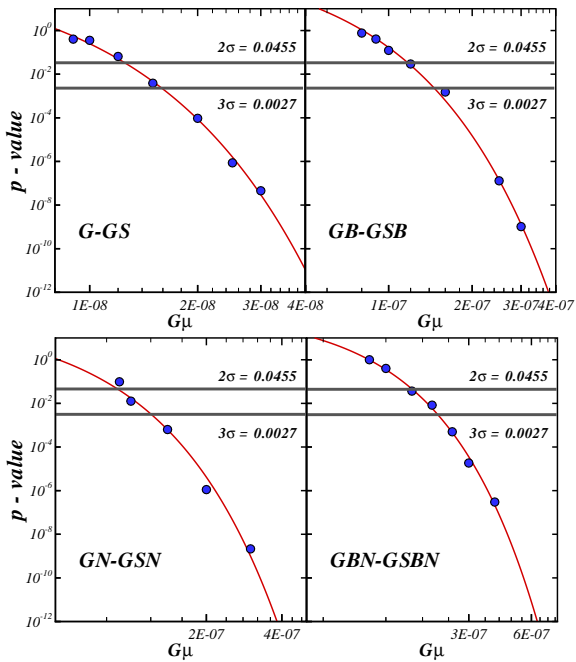} \\
\caption {\small P-value as a function of $G\mu$. Upper left panel indicates the result for Gaussian CMB map. The effect of beam on the capability of TPCF to detect CS has been shown in upper right panel. Lower left panel corresponds to P-value for map in the presence of instrumental noise. The effect of finite beam size and instrumental noise have been indicated in lower right panel. To determine each point in this plot we did average on at least 100 ensembles.}\label{pvalue1}
\end{figure}

We then define $\chi^{2} \equiv -2\sum \ln p(\theta)$. The final P-value related to $\chi^{2}$ is calculated based on chi-square distribution function with
 $2(\theta_{max}-\theta_{min})/\Delta \theta -2$ 
degrees of freedom.  Fig. \ref{pvalue1} shows this P-value as a function of $G\mu$ for various maps with $\Theta=10^\circ$.  We have drawn lines at $p=0.0027$, and $0.0455$, since these correspond to 3$\sigma$ and 2$\sigma$ significance levels.  This shows that the TPCF can detect CS at $95\%$CL provided $G\mu \gtrsim 1.2 \times 10^{-8}$ in maps without instrumental noise. If noise is present, with rms $\sigma_{noise}=10\mu$K, then this limit increases to $G\mu \gtrsim 9.0 \times 10^{-8}$.  Including beam smearing further degrades our limits:  the minimum detectable CS becomes $G\mu\gtrsim 1.6 \times 10^{-7}$ at $2\sigma$ confidence interval.  Table \ref{tab1} summarizes our results.

To minimize the effect of cosmic variance more than previous task, we also construct new quantities \citep{Rossi11}:
\begin{equation}
\bar{\xi}_{(\diamond)}(\theta)\equiv \frac{\xi_{(\diamond)}(\theta)-\xi_{(\otimes)}(\theta)}{\xi_{(\otimes)}(\theta)}
\end{equation}
 and again compute 
\begin{equation}\label{t-test1}
\bar{t}(\theta)=\frac{\bar{\xi}_{(\diamond)}(\theta)}{\sqrt{\bar{\sigma}_{(\diamond)}^{2}(\theta)}}
\end{equation}
 The $\chi^2$ for the optimum interval of $\theta$, for the p-value of $\bar{t}(\theta)$ has been computed. Then we detremined the final p-value for computed $\chi^2$.
 Our results are in agreement with previous upper bound on $G\mu$ mentioned in Table \ref{tab1}.

\section{Conclusion}
If they exist, cosmic strings are expected to leave an imprint in the CMB.  We argued that although such strings may alter the power spectrum (Fig. \ref{power1}) and the statistics of hot and cold spots (Figs. \ref{map2} and \ref{numberdensity2}), the change to the one-point stastistics of peaks in CMB maps cannot be distinguished from that for a Gaussian field having the same power spectrum (Fig. \ref{numberdensity}).  On the other hand, the two-point statistics show differences (Figs.~\ref{gs-gaussian} and \ref{two22}) which we believe can be used to reject the hypothesis that the map is a purely Gaussian (Fig.~\ref{pvalue1}).  

We argued that CS will be detected at high significance only if the string tension is sufficiently high:  $G\mu \gtrsim 1.2\times 10^{-8}$.  Accounting for the fact that instrumental noise complicates the measurement increases this limit to $G\mu \gtrsim 9.0\times 10^{-8}$ (Fig. \ref{pvalue1} and Table \ref{tab1}).  The CS signal is particularly strong on arcminute and smaller scales.  Some of this signal is removed if the beam size of the experiment is larger than this scale.  For a $4'$ beam, the limit is $G\mu  \gtrsim 1.2 \times 10^{-7}$ at $2\sigma$ confidence interval.
Broader beams further degrade the limit on $G\mu$.


\begin{table}
\begin{center}
\begin{tabular}{|c|c|c|}
\hline  Map & $2\sigma$ & $3\sigma$   \\
\hline  G-GS  & $G\mu\gtrsim 1.2\times 10^{-8}$  & $G\mu\gtrsim 1.6\times 10^{-8}$ \\
\hline  GN-GSN&$G\mu\gtrsim 9.0\times 10^{-8}$ & $G\mu\gtrsim 1.2\times 10^{-7}$ \\
\hline  GB-GSB& $G\mu\gtrsim 1.2\times 10^{-7}$ & $G\mu\gtrsim 1.5\times 10^{-7}$ \\
\hline  GBN-GSBN&$G\mu\gtrsim 1.6\times 10^{-7}$  & $G\mu\gtrsim 2.2 \times 10^{-7}$ \\
\hline
   \end{tabular}
\end{center}
\caption{\label{tab1} Limits on $G\mu$ which come from analysis of the differences between the TPCF for maps with and without cosmic strings, when $R=1'$, FWHM=$4'$ and $\sigma_{noise}=10\mu K$.}
\end{table}

We have argued that two-point statistics of peaks (the pair correlation function) are better than one-point statistics (peak number counts) for distinguishing between models.  Our results suggest that the $n$-point correlation functions {\em of peaks} can be used for similar purpose.  This is interesting in view of previous work showing that the 3-point statistics of all pixels is not very informative.  

Final remark is that it could be interesting to use more realistic models
\citep{land03,fra08,hind09,hind10,shellard10} to simulate a map taking all
contributions of cosmic strings into account and apply our method to examine  the effect of
cosmic strings in our future works. 

\section*{Acknowledgments}
MSM is grateful to the Office of Associates at ICTP and the hospitality of HECAP section of ICTP.  MSM and BJ are grateful to  H. Moshafi for preparing some power spectra. BJ thanks M. Yazdizadeh for his help with computing. We acknowledge the use of CAMB software, WMAP-7, the SNLS gold sample and the Two Degree Field Galaxy Redshift Survey (2dFGRS) data sets. We also thank to anonymous referee to help us to improve the manuscript.  MSM thanks the Shahid Beheshti University research deputy affairs which supported this work by  grant No. 600/1037.


\end{document}